\documentclass[CEJP,PDF]{cej} 
\usepackage{layout}
\usepackage{amsmath}
\usepackage{textcomp}
\usepackage{hyperref}

\title{Probing the QCD Critical Point with Relativistic Heavy-Ion Collisions}

\articletype{Research Article}

\author{Steffen A. Bass\inst{1}\email{bass@phy.duke.edu}, Hannah Petersen\inst{1},
        Cory Quammen\inst{2}, Hal Canary\inst{2}, 
         Christopher G. Healey\inst{3}
        and Russell M. Taylor II\inst{2}}

\institute{
     \inst{1} Department of Physics, Duke University\\
     Durham, NC 27708-0305, USA
     \inst{2} Department of Computer Science, University of North Carolina\\
     Chapel Hill, NC 27599-3175, USA
     \inst{3} Department of Computer Science, North Carolina State University\\
     Raleigh, NC 27695-8206, USA
          }

\abstract{We utilize an event-by-event relativistic hydrodynamic
calculation performed at a number
of different incident beam energies
to investigate the creation of hot and dense QCD matter near the critical point. 
Using state-of-the-art analysis and visualization tools we demonstrate that each
collision event probes QCD matter characterized by a wide range of temperatures and 
baryo-chemical potentials, making a dynamical response of the system to the vicinity
of the critical point very difficult to isolate above the background.}


\begin{document}
\maketitle


\section{Introduction}

Over the past decade, significant progress has been made in our understanding
of the QCD equation of state, in particular at small values of the baryo-chemical
potential \cite{Aoki:2006we,Bazavov:2009zn,Borsanyi:2010cj}. This progress has been due to improved Lattice Gauge Theory calculations 
as well as due to high quality data from RHIC and LHC that can be connected to the
Lattice calculations via relativistic fluid dynamical simulations \cite{Huovinen:2009yb,Song:2010mg}. With new data
from the RHIC beam-energy scan and the start of construction of the FAIR project, renewed
attention has been cast on the QCD equation of state at non-vanishing values of the 
baryo-chemical potential. Among the most intriguing features of the QCD equation of
state in that regime is the possible existence of a critical point \cite{Stephanov:1998dy,Fodor:2001pe,Athanasiou:2010kw}. While current
theoretical efforts in determining the position of the critical point -- or even it's 
mere existence -- are hampered by large systematic uncertainties, the presence of
the critical point may affect
the dynamics of heavy-ion collisions and manifest itself in measurable quantities, thus
shedding light on its location from an experimental point of view.

A necessary prerequisite for any measurable response of the collision system to
the presence of the critical point is that a sufficient volume of the created QCD matter
at some point during its evolution attains values of the temperature and chemical
potential within the critical region of $T$ and $\mu_B$ values in which the properties
of matter would be (significantly) modified due to the presence of the critical point.
Whether or not satisfying this prerequisite will actually allow for an experimental
estimate of the location of the critical point will depend on multiple other conditions
such as
the strength of the medium modification due to the critical point, the influence of the
subsequent evolution in the hadronic phase on the observables, the width of the 
critical region and the range of temperatures and chemical potentials probed in a
single collision at a given beam energy. In this paper, we shall focus on the latter
point, namely conduct an analysis on the range of temperatures and chemical potentials
that are probed at any given time in a single heavy-ion collision.

\section{Modeling of Heavy-Ion Collisions}
For our analysis, we utilize a state-of-the-art (hybrid) hydrodynamic model with
fluctuating initial conditions \cite{Petersen:2008dd}. The initial conditions for the event-by-event hybrid approach are generated by the Ultra-relativistic Quantum Molecular Dynamics 
approach (UrQMD) \cite{Bass:1998ca,Bleicher:1999xi}. The nucleon distributions in the two nuclei are sampled according to Wood-Saxon profiles and the
subsequent interactions among the nuclei as well as secondary produced particles are calculated in the hadron transport approach and include the excitation and fragmentation of strings as well as resonance dynamics . At that point in time when the two nuclei have passed through each other - this time depends on the beam energy of the collision - all particles are converted to an energy density distribution that serve as initial condition for the hydrodynamic expansion of the quark gluon plasma. 

\begin{figure}
\includegraphics[width=0.95\textwidth]{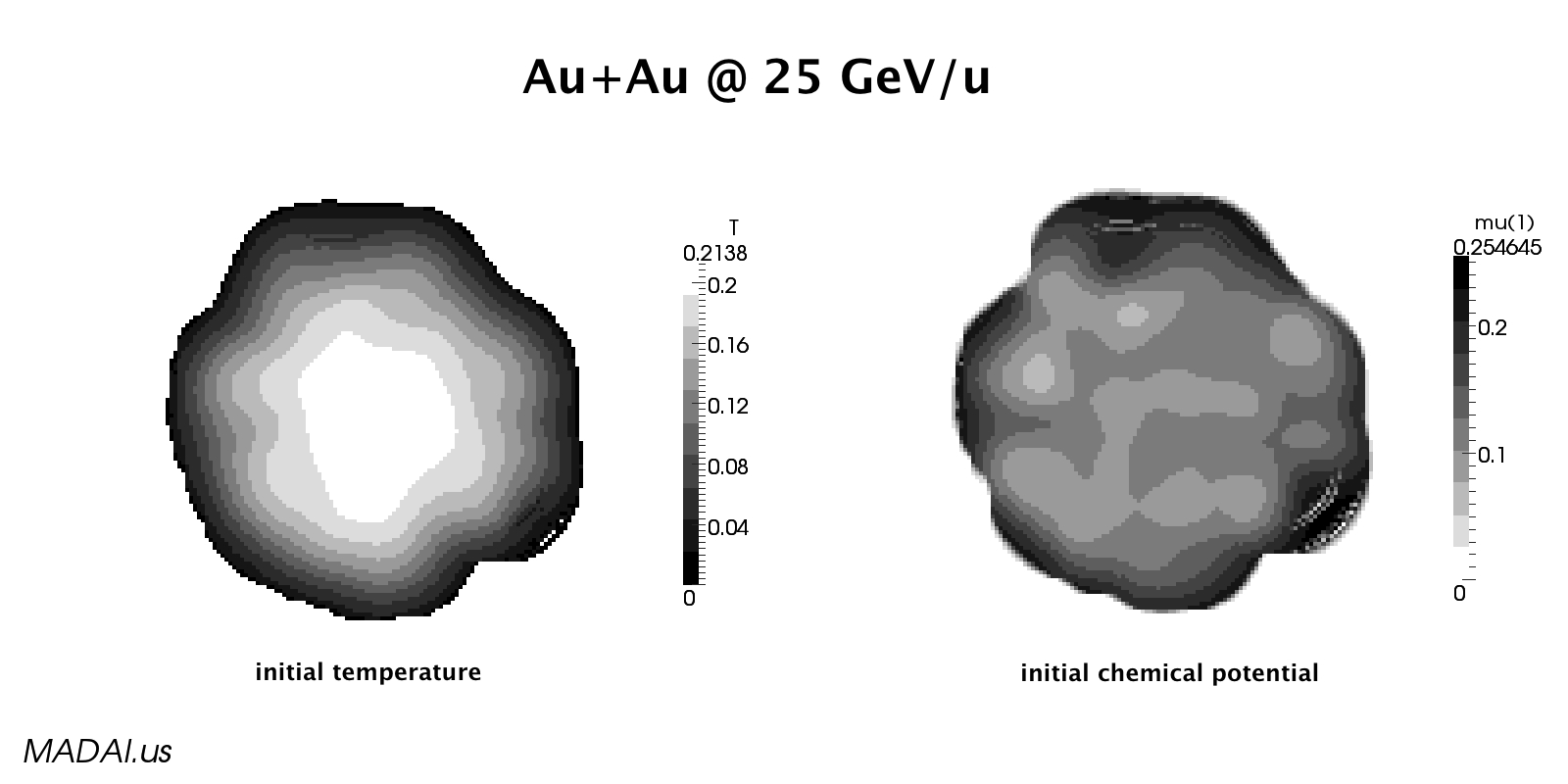}
\caption{Distribution of temperatures (left) and quark-chemical potential
$\mu_q$ (right) for a single event initial condition at 25 GeV/u beam energy.\label{fig1}}
\end{figure}

Naturally, these fluctuating initial conditions span a wide range in temperature and baryo- or quark-chemical potential as shown in Fig. \ref{fig1}. At a beam energy of $E_{\rm lab}=25A$ GeV that is foreseen as an energy reached by the future FAIR-SIS300 accelerator, the initial temperatures in the central transverse slice range from 120-210 MeV. In the contour plot of the baryo-chemical potential even more granular structures are visible. This is a more realistic treatment than in the smooth averaged profiles  that have been used in many hydrodynamical calculations before. 


The equation of state (EoS) used for the hydrodynamic stage of our calculation incorporates 
a critical end point (CEP) in line with lattice data. The EoS has been constructed
by coupling the Polyakov loop (as an order parameter for deconfinement) to a chiral hadronic $SU(3)_f$ model. In this configuration the EoS describes chiral restoration as well as the deconfinement phase transition (for more details on the EoS we refer to \cite{Steinheimer:2007iy}).

The hybrid approach includes also a treatment of the hadronic phase by feeding the particles back into the hadron transport approach after the freeze-out criterion in the hydrodynamic calculation has been reached. This is important for the quantitative prediction of observables that are sensitive to the critical point. In this work, we concentrate on pointing out the spread in the phase-diagram of single collisions and therefore do not show the subsequent hadronic evolution. 

\section{Results and Discussion}

\begin{figure}[t]
\includegraphics[width=0.95\textwidth]{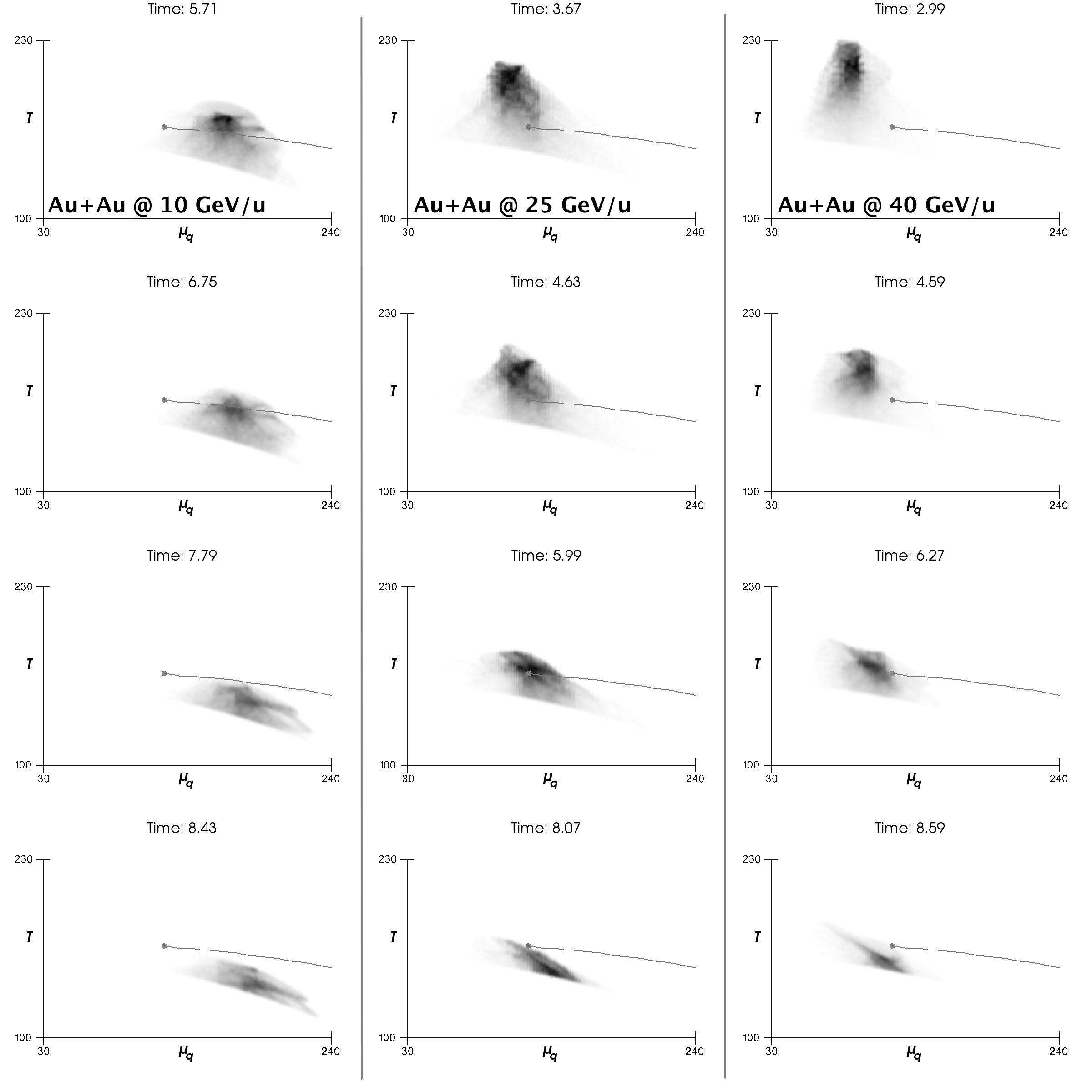}
\caption{Time evolution (top to bottom) of QCD matter created in central Au+Au
collisions at 10 (left), 25 (center) and 40 (right) GeV/u beam energy projected into
the $T-\mu_q$ phase diagram. The grey shading represents the amount of matter present
at the respective value of $T$ and $\mu_q$. \label{fig2}}
\end{figure}

In our analysis we focus on the time-evolution of the hot and dense QCD medium in
$T-\mu$ space to highlight the range of temperatures and baryo-chemical potentials
present in the medium at any given time during the evolution. Figure~\ref{fig2} shows
the analysis for three individual events at three different beam energies, 
10, 25 and 40 GeV/u (in the left, 
center and right columns respectively) that are representative 
of the range of available beam energies at current and future facilities at RHIC and FAIR.
The abscissa denotes the quark chemical potential in MeV and the ordinate shows the
temperature (also in MeV). As can be clearly seen, a realistic event will not follow
a single sharp trajectory of constant entropy per baryon in $T-\mu_q$ space, but will
cover a wide range of $T,\mu_q$ values at any given time.
For each event we have selected four time-steps at the early,
intermediate and later stages of the evolution, with particular emphasis on capturing
at least one time-step that is close to the CEP. The analysis has been performed
utilizing the {\sc MADAI Workbench} visualization and analysis package \cite{madaiworkbench}, 
that is based on
{\sc Paraview} \cite{paraview} but contains additional tools, e.g. a Gaussian splatter filter
that enables the use of the continuous shading seen in the figure with darker shades
of grey representing a higher concentration of QCD matter at that particular point
in the $T-\mu$ diagram. We find that for the particular EoS used
in this calculation, an incident beam energy of 25 GeV/u seems to provide the best 
opportunity for creating and probing QCD matter in the vicinity of the CEP. This 
beam energy should be easily accessible in a systematic way with the FAIR facility at GSI.

While Figure~\ref{fig2} does make a promising case for collisions at $\sim 25$ GeV/u
probing the CEP, it also clearly shows the challenges that an experimental
determination of the CEP would face: each event contains QCD matter spread over 
at least 50~MeV in $\mu_q$ with the three beam energies being separated by a similar
difference in terms of $\mu_q$. Presumably, the largest response of the system would
be obtainable if the critical range around the CEP were of similar width in $\mu_q$ as
the event itself; for a range significantly smaller than that, matter outside the
critical range may very well overwhelm any signal stemming from matter passing through
the critical region.

\section{Summary and Outlook}

We have shown that event-by-event hydrodynamic modeling of heavy-ion collisions 
creates QCD matter 
at a  broad range of $T$ and $\mu_q$ values for any given time during
the evolution of the system, significantly complicating the possible response  of 
the system to the presence 
of a CEP in the QCD phase-diagram via experimental/phenomenological methods. For the
EoS used in our calculation, an incident beam energy of 25 GeV/u seems to provide the best 
opportunity for creating and probing QCD matter in the vicinity of the CEP.
Any future realistic calculation of observable measures for the presence of a CEP will have
to take multiple features of the systems's evolution into account, including the 
aforementioned range of $T,\mu_q$ values probed, the width of the critical region 
as well as the influence of the hadronic phase. We hope to address some of 
these additional items in the near future.

\section*{Acknowledgements}

H.P. acknowledges a Feodor Lynen
fellowship of the Alexander von Humboldt
foundation. This work was supported in part by U.S. department of Energy grant
DE-FG02-05ER41367 and NSF grant PHY-09-41373. 

\bibliographystyle{h-physrev}
\bibliography{/Users/bass/Publications/SABrefs}


\end{document}